\documentclass{article}
\usepackage{spconf,amsmath,graphicx}
\pdfoutput=1

\usepackage{tikz}
\usepackage{amssymb}
\usepackage{algorithm,setspace}
\usepackage[noend]{algpseudocode}

\usepackage{pgfplots}
    \usetikzlibrary{calc,positioning,patterns}
    \pgfplotsset{compat=1.3}
\usepackage{pgfplotstable}

\newcommand{\C}[0]{\mathbb{C}}

\newcommand{\x}[1]{\boldsymbol{{#1}}_{t, f}}
\newcommand{\est}[1]{\widehat{\boldsymbol{{#1}}}_{t, f}}

\tikzstyle{block} = [draw, fill=blue!14, rectangle, minimum height=3em, minimum width=5em, align=center]
\tikzstyle{sum} = [draw, fill=white, circle, node distance=1cm]
\tikzstyle{input} = [coordinate]
\tikzstyle{output} = [coordinate]
\tikzstyle{pinstyle} = [pin edge={to-,thin,black}]
\tikzstyle{branch}=[fill,shape=circle,minimum size=3pt,inner sep=0pt]
\tikzstyle{connarrow}=[-latex, line width=1pt]
\tikzstyle{connline}=[-, line width=1pt]

\title{Customizable end-to-end optimization of online neural network-supported dereverberation for hearing devices}
\name
 {Jean-Marie Lemercier$^{\star}$, Joachim Thiemann$^{\dagger}$, Raphael Koning$^{\dagger}$, Timo Gerkmann$^{\star}$\thanks{This work has been funded by the Federal Ministry for Economic Affairs and Climate Action, project 01MK20012S, AP380. The authors are responsible for the content of this paper.}}
	\address{$^{\star}$Signal Processing (SP), Universität Hamburg, Germany\\ 
	$^{\dagger}$Advanced Bionics, Hanover, Germany\\
	{\tt \small \{firstname.lastname\}@uni-hamburg.de, \{firstname.lastname\}@advancedbionics.com}
	}
	
\begin{document}
\ninept
\maketitle
\begin{abstract}

This work focuses on online dereverberation for hearing devices using the weighted prediction error (WPE) algorithm. WPE filtering requires an estimate of the target speech power spectral density (PSD). Recently deep neural networks (DNNs) have been used for this task. However, these approaches optimize the PSD estimate which only indirectly affects the WPE output, thus potentially resulting in limited dereverberation. 
In this paper, we propose an end-to-end approach specialized for online processing, that directly optimizes the dereverberated output signal. In addition, we propose to adapt it to the needs of different types of hearing-device users by modifying the optimization target as well as the WPE algorithm characteristics used in training. 
We show that the proposed end-to-end approach outperforms the traditional and conventional DNN-supported WPEs on a noise-free version of the WHAMR! dataset.

\end{abstract}
\begin{keywords}
online algorithm, dereverberation, neural network, end-to-end learning, hearing devices
\end{keywords}
\section{Introduction}
\label{sec:intro}

Communication and hearing devices require modules aiming at suppressing undesired parts of the signal to improve the speech quality and intelligibility. Reverberation is one of such distortions caused by room acoustics, and is characterized by multiple reflections on the room enclosures. Late reflections particularly degrade the speech signal and may result in a reduced intelligibility \cite{Kuttruff2016,Naylor2011}.

Many traditional approaches were proposed for dereverberation such as spectral enhancement \cite{Habets2007}, beamforming \cite{Kuklasinski2015}, a combination of both \cite{Cauchi2015}, coherence weighting \cite{Allen1977, Schwarz2015, Gerkmann2011}, and linear-prediction based approaches such as the weighted-prediction error (WPE) algorithm \cite{Nakatani2008b,Jukic2015}. WPE computes an auto-regressive multi-channel filter and applies it to a delayed group of reverberant speech frames. The approach is able to cancel late reverberation while preserving early reflections, thus improving speech intelligibility for normal and hearing-aided listeners \cite{Warzybok2003,Bradley2003}. WPE and its extensions have been shown to be robust and efficient multi-channel techniques. However, these methods require the prior estimation of the anechoic speech power spectrum density (PSD), which is modelled for instance through the speech periodogram \cite{Nakatani2008b}, by an autoregressive process \cite{Yoshioka2011} or through non-negative matrix factorization \cite{Kagami2018}. A deep neural network (DNN) was first proposed in \cite{Kinoshita2017} to model the anechoic PSD, thus avoiding the use of an iterative refinement process.

As hearing devices require to operate in real-time in variable environments, the methods implemented should be suited for frame-to-frame online processing, as well as being adaptive to changing room acoustics. Online adaptive approaches are based on either Kalman filtering \cite{Schwartz2015, Braun2016} or on a recursive least squares (RLS) adapted WPE. In this latter RLS-WPE framework, the PSD is either estimated by recursive smoothing of the reverberant signal \cite{Caroselli2017} or by a DNN \cite{Heymann2018}.

In the previously cited work, the neural network was trained toward PSD estimation, although the aim of the algorithm is WPE-based dereverberation. End-to-end techniques were proposed, using an Automatic Seech Recognition (ASR) criterion in order to refine the front-end DNN handling e.g. speech separation \cite{Chang2019}, denoising \cite{Ochiai2017}, or multiple tasks \cite{Zhang2021}. An end-to-end procedure for online dereverberation and ASR based on DNN-WPE was proposed in \cite{Heymann2019}. However, for hearing devices, it is less clear which criterion reaches optimal speech intelligibility and quality, and such performance is highly dependent on the considered user category.

In this work, we propose to use a criterion on the WPE output short-time spectrum for online dereverberation to improve instrumentally predicted speech intelligibility and quality. To solve the issue of the initialization period of RLS-WPE, we design a dedicated training procedure taking into account the adaptive nature of the algorithm. Finally we include a specialization toward different hearing-device users categories: hearing-aid  (HA) users on the one hand benefiting from early reflections like normal listeners \cite{Warzybok2003}; cochlear-implanted (CI) on the other hand which do not benefit from early reflections \cite{Hu2014}.

The rest of this paper is organized as follows.  In Section \ref{sec:WPE}, the online DNN-WPE dereverberation scheme is summarized, followed by a description of the proposed end-to-end training procedure in Section \ref{sec:e2e}. The experimental setup is described in Section \ref{sec:majhead} and the evaluation results are presented and discussed in Section \ref{sec:results}.

\section{Signal model and DNN-supported WPE Dereverberation}
\label{sec:WPE}

\subsection{Signal model}

In the short-time Fourier transform (STFT) domain using the subband-filtering approximation \cite{Nakatani2008b}, the reverberant speech $\boldsymbol{x} \in \C^{D}$ is obtained at the $D$-microphone array by convolution of the anechoic speech $s$ and the room impulse responses (RIRs) $\boldsymbol{H} \in \C^{D \times D}$ with length $L$,

\begin{equation} \label{eq:signal_model0}
    \x{x} = \displaystyle{\sum_{\tau = 0 }^{L}}\boldsymbol{H}_{\tau, f} s_{t-\tau, f} = \x{d} + \x{e} + \x{r} ,
\end{equation}

\noindent where $t$ denotes the time frame index and $f$ the frequency bin, which we will drop when not needed.
$\boldsymbol{d}$ denotes the direct path, $\boldsymbol{e}$ the early reflections component, and  $\boldsymbol{r}$ the late reverberation. The early reflections component $\boldsymbol{e}$ was shown to contribute to speech quality and intelligibility for normal and HA listeners \cite{Bradley2003} but not for CI listeners, particularly in highly-reverberant scenarios \cite{Hu2014}. Therefore, we propose that the dereverberation objective is to retrieve $\boldsymbol{\nu} = \boldsymbol{d + e}$ for HA listeners and $\boldsymbol{\nu} = \boldsymbol{d}$ for CI listeners.

\subsection{WPE dereverberation}

In relation to the subband reverberant model in \eqref{eq:signal_model0}, the WPE algorithm \cite{Nakatani2008b} uses an auto-regressive model to approximate the late reverberation $\boldsymbol{r}$. Based on a zero-mean time-varying Gaussian model on the STFT anechoic speech $s$ with time-frequency dependent PSD $\lambda_{t,f}$, a multi-channel filter $\boldsymbol{G} \in \C^{DK \times D}$ with $K$ taps is estimated. This filter aims at representing the inverse of the late tail of the RIRs $\boldsymbol{H}$, such that the target $\boldsymbol{\nu}$ can be obtained through linear prediction, with a delay $\Delta$ avoiding undesired short-time speech cancellations, which also leads to preserving parts of the early reflections:

\begin{equation} \label{eq:MCDLP}
    \est{\nu} = \boldsymbol{x}_{t, f} -   \boldsymbol{G}^H_{t,f} \boldsymbol{X}_{t - \Delta, f} ,
\end{equation}
where $\boldsymbol{X}_{t - \Delta, f} = \begin{bmatrix} \boldsymbol{x}^T_{t-\Delta, f}, \dots, \boldsymbol{x}^T_{t-\Delta-K+1, f} \end{bmatrix}^T \in \C^{DK}$.

\vspace{0.2em}
In order to obtain an adaptive and real-time capable approach, RLS-WPE was proposed in \cite{Caroselli2017}, where the WPE filter $\boldsymbol{G}$ is recursively updated along time:

\begin{equation} \label{eq:kalman} \x{K} = \displaystyle{\frac{
            (1 - \alpha) \boldsymbol{R}^{-1}_{t-1, f} \boldsymbol{X}_{t - \Delta, f} }{
            \alpha \lambda_{t, f} + (1 - \alpha) \boldsymbol{X}^H_{t - \Delta, f} \boldsymbol{R}^{-1}_{t-1, f} \boldsymbol{X}_{t - \Delta, f} }} , \end{equation}
            
\begin{equation} \label{eq:invcov} \boldsymbol{R}^{-1}_{t, f} = \frac{1}{\alpha} \boldsymbol{R}^{-1}_{t-1, f} -  \frac{1}{\alpha} \x{K} \boldsymbol{X}^T_{t-\Delta, f} \boldsymbol{R}^{-1}_{t-1, f} , \end{equation}
            
\begin{equation} \label{eq:filter} \boldsymbol{G}_{t, f} = \boldsymbol{G}_{t-1, f}  + \x{K} (\boldsymbol{x}_{t, f} - \boldsymbol{G}_{t-1, f}^H  \boldsymbol{X}_{t-\Delta, f})^H. \end{equation}

 \noindent$\boldsymbol{K} \in \C^{DK}$ is the Kalman gain, $\boldsymbol{R} \in \C^{DK \times DK}$ the covariance of the delayed reverberant signal buffer $\boldsymbol{X}_{t - \Delta, f}$ weighted by the PSD $\lambda$, and $\alpha$ the forgetting factor. 

\subsection{DNN-based PSD estimation}

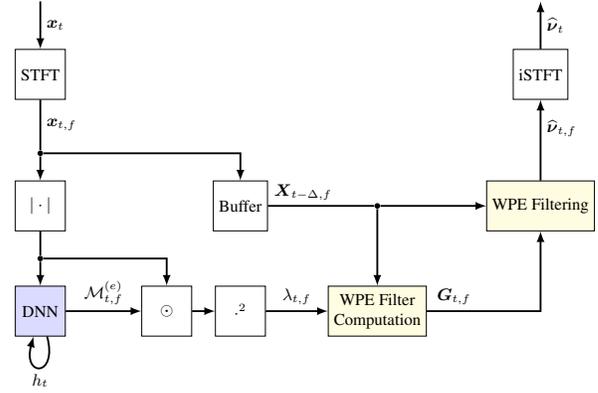
\begin{figure}
    \centering
    \tikzstyle{block} = [draw, rectangle, minimum height=3em, minimum width=3em]
    \scalebox{0.7}{
    \begin{tikzpicture}[auto]
    
        \node at (-5.0, 0.0) (input) {};
        \node[block] at (-5.0, -1.5) (stft) {STFT};
        \node[branch] at (-5.0, -3.0) (branch_in_buffer) {};
        \node[block] at (-1.2, -4.0) (buffer) {Buffer};
        \node[branch] at (1.4, -4.0) (bufferbranchout) {};

        \node[block] at (-5.0, -4.0) (mag) {$| \cdot |$};
        \node[branch] at (-5.0, -5.0) (branch_in_mask) {};

        \node[block, fill=blue!14] at (-5.0, -6.0) (lstm) {DNN};
        
        \node[block] at (-2.6, -6.0) (mask) {$\odot$};
        \node[block] at (-1.2, -6.0) (square) {$\cdot^2$};

        \node[block, fill=yellow!14, align=center] at (1.4, -6.0) (wpe) {WPE Filter\\Computation};
        
        \node[block, fill=yellow!14] at (4.5, -4.0) (filter) {WPE Filtering};
        
        \node[block] at (4.5, -1.5) (istft) {iSTFT};

        \node at (4.5, 0.0) (output) {};

        \draw[connarrow] (input) -- node[pos=0.5] {$\boldsymbol{x}_t$} (stft);
        \draw[connline] (stft) -- node[pos=0.5] {$\boldsymbol{x}_{t,f}$} (branch_in_buffer);
        \draw[connarrow] (branch_in_buffer) -| (buffer);
        \draw[connarrow] (branch_in_buffer) -- (mag);
        
        \draw[connline] (mag) -- (branch_in_mask);
        \draw[connarrow] (branch_in_mask) -- (lstm);
        
        \draw[connarrow] (branch_in_mask) -| (mask);
        \draw[connarrow] (lstm) -- node[pos=0.5] {$\mathcal{M}^{(e)}_{t,f}$} (mask);
        \draw[connarrow] (lstm) to[loop below] node[pos=0.5] {$h_t$} (lstm);

        \draw[connarrow] (mask) -- (square);
        \draw[connarrow] (square) -- node[pos=0.5] {$\lambda_{t,f}$} (wpe);
        
        \draw[connline] (buffer)  -- node[pos=0.33]  {$\boldsymbol{X}_{t-\Delta,f}$} (bufferbranchout);
        \draw[connarrow] (bufferbranchout) -- (wpe);
        \draw[connarrow] (bufferbranchout) -- (filter);
        \draw[connarrow] (wpe) -| node[pos=0.12,above] {$\boldsymbol{G}_{t,f}$} (filter);
        
        \draw[connarrow] (filter) -- node[pos=0.65,right] {$\est{\nu}$}  (istft);
        \draw[connarrow] (istft) -- node[pos=0.5,right] {$\widehat{\boldsymbol{\nu}}_t$}  (output);

    \end{tikzpicture}
    }
    \caption{\protect\centering DNN-supported online WPE dereverberation. Blue blocks refer to trainable neural network layers. Yellow blocks represents adaptive statistical signal processing}
    \label{fig:dnn-wpe}
\end{figure}

The anechoic speech PSD $\lambda_{t,f}$ is estimated at each time step $t$, either by recursive smoothing of the reverberant periodogram \cite{Caroselli2017} or with help of a DNN \cite{Heymann2018}.  A block diagram of the DNN-WPE algorithm as proposed in \cite{Heymann2018} is given in \figurename~\ref{fig:dnn-wpe}. In this approach, the channel-averaged magnitude frame $|\boldsymbol{\bar{x}}_t|$ is fed as input to a recurrent neural network with state $h_t$ and the output is a target speech mask $ \mathcal{M}^{(\nu)}_{t,f}$. The PSD estimate is then obtained by time-frequency masking:
\begin{equation} \label{eq:lambda}
\widehat{\lambda}_{t, f} = ( \mathcal{M}^{(\nu)}_{t, f} \odot |\boldsymbol{\bar{x}}_{t, f}| )^2.
\end{equation}

The DNN is optimized with a mean-squared error criterion on the masked output in \cite{Kinoshita2017, Heymann2018}. In contrast, we propose to use the Kullback-Leibler (KL) divergence as it led to better results:

\begin{equation}  \label{eq:loss-nwpe}
     \mathcal{L}_\text{DNN-WPE} = \text{KL} ( \mathcal{M}^{(\nu)}_{t,f} \odot |\boldsymbol{\bar{x}}_{t, f} | , | \x{\nu} | ).
\end{equation}

The training objective $\mathcal{L}_\text{DNN-WPE}$ does not match the output $\widehat{\boldsymbol{\nu}}$ of the whole algorithm, thus potentially limiting the dereverberation performance.

\section{Proposed End-to-End Training Procedure for Online Dereverberation Optimality}
\label{sec:e2e}

\subsection{End-to-end criterion and objectives}

Here we propose an end-to-end training procedure where the optimization criterion is placed at the output of the DNN-WPE algorithm. The objective is to include the back-end WPE into the computations through which the loss will be backpropagated during training:

\begin{equation} \label{eq:loss-e2e}
    \mathcal{L}_\text{E2E} = \text{KL}( | \est{\nu} | , | \x{\nu} | ).
\end{equation}

In contrast to \cite{Heymann2019}, no ASR criterion is used here. Instead, the loss is computed in the time-frequency domain. This enables us to take different targets and WPE parameters into consideration, for customizing the approach towards different hearing-device user categories. Namely, for HA listeners, where early reflections are considered beneficial \cite{Bradley2003}, we set the training target to $\boldsymbol{\nu} = \boldsymbol{d}+\boldsymbol {e}$ and we use a larger prediction delay $\Delta_{\text{HA}}$. For CI listeners, for which early reflections may be harmful \cite{Hu2014}, we set $\boldsymbol{\nu} = \boldsymbol{d}$ and we use a shorter delay $\Delta_{\text{CI}}<\Delta_{\text{HA}}$ to remove as much of the early component as possible given the delayed linear prediction model \eqref{eq:filter}.

\subsection{Initialization period}

As all operations in RLS-WPE are differentiable, we can use backpropagation through the whole WPE algorithm. However, an important practical aspect of this study focuses on handling the initialization period of the RLS-WPE algorithm. During this interval of $L$ time frames, the filter $\boldsymbol{G}$ has not yet converged to a stable value, and the resulting dereverberation performance is suboptimal, as we will show it in the experiments (see Section 5).
 
Therefore, rather than relying on a hypothetical shortening of this period through implicit PSD optimization \cite{Heymann2019}, we choose to exclude this initialization period from training, which leads us to design the procedure as given in Algorithm \ref{alg:training}. 
Finally, we investigate using a pretrained DNN, trained on the same dataset with the loss function \eqref{eq:loss-nwpe}, and plugging it into Algorithm \ref{alg:training} for fine-tuning.

\begin{algorithm}[h]
\caption{End-to-End Training Procedure}\label{alg:training}

\begin{algorithmic}[1]
\setstretch{1.1}
\State Extract STFT of given sequence
\State Segment sequence in $N$ segments of size $L$
\For{$n \in \{0 \dots N-1 \}$}

\hrulefill
\If{$n = 0$}
\Comment{Initialization period}
\State Initialize LSTM state $h^{(0)}_0 = 0$
\State Initialize WPE statistics 
\vspace{-0.7em} \begin{equation*} \boldsymbol{G}^{(0)}_{0, f} = \boldsymbol{0} \, , \, ( \boldsymbol{R}^{-1} )^{(0)}_{0, f} = \boldsymbol{I} \end{equation*}
\itemsep-0.7em
\For{$t \in \{ 0 \dots L-1 \} $}
\itemsep0em
\State Compute $\est{e}$ with one pass of DNN-WPE
\EndFor
\EndIf

\hrulefill
\If{$n > 0$}
\Comment{After initialization}
\State Initialize LSTM state $h^{(n)}_0 = h^{(n-1)}_{L-1}$
\State Initialize WPE statistics
\vspace{-0.7em} \begin{equation*} \boldsymbol{G}^{(n)}_{0, f} = \boldsymbol{G}^{(n-1)}_{L-1, f} \, , \, ( \boldsymbol{R}^{-1} )^{(n)}_{0, f} = ( \boldsymbol{R}^{-1} )^{(n-1)}_{L-1, f} \end{equation*}
\itemsep-0.7em
\For{$t \in \{ 0 \dots L-1 \} $}
\itemsep0em
\State Compute $\est{e}$ with one pass of DNN-WPE
\EndFor
\State Backpropagate loss \eqref{eq:loss-e2e} through time on $n$
\State Repeat [13:] to re-update $h^{(n)}_{L-1}, \boldsymbol{G}^{(n)}_{L-1, f}$
\EndIf

\EndFor

\end{algorithmic}
\end{algorithm}

\section{Experimental Setup}
\label{sec:majhead}

\subsection{Dataset generation}

The data generation is inspired from the WHAMR! dataset \cite{Maciejewski2020} and uses anechoic speech utterances from the WSJ0 dataset. As the initialization time $L$ typically corresponds to $4$ seconds when using a forgetting factor of $\alpha = 0.99$, we concatenate utterances belonging to the same speaker and construct sequences of approximately $20$ seconds. Within each sequence, permutations of the utterances are used to create several versions of the sequence, so as not to lose too much data since the first segment is never used for optimization.

These sequences are convolved with $2$-channel RIRs generated with the RAZR engine \cite{Wendt2014a} and randomly picked. Each RIR is generated by uniformly sampling room acoustics parameters as in \cite{Maciejewski2020} and a $\text{T}_\text{60}$ reverberation time between $0.4$ and $1.0$ seconds. As target data for the HA case, the first $40$~ms of the RIR is convolved with the utterance, representing the direct path and the early reflections, whereas for the CI scenario, only the direct path is retained. Each training set consists of approximately $55$ hours of speech data sampled at $16$~kHz. 

\subsection{Hyperparameter settings}

All approaches are trained by backpropagating the KL divergence through time, using the Adam optimizer with a learning rate of $10^{-4}$, exponentially decreasing by a factor of $0.96$ at every epoch. Early stopping with a patience of $10$ epochs and mini-batches size of $128$ segments are used. The STFT uses a square-rooted Hann window of $32$~ms and a $75$~\% overlap, and segments of $L=4$~s are constructed.

The WPE filter length is set to $K=10$ STFT frames ($\sim 80\mathrm{ms}$) as our goal is to focus on the beginning of the reverberation tail, where most of the reverberant energy lies.
Another reason is that the WPE computational complexity globally increases with the square of $K$, making end-to-end training longer and more unstable.

The number of channels is $D=2$, the adaptation factor $\alpha=0.99$ and the delays $\Delta_{\text{HA}}=5$ frames for the HA scenario and $\Delta_{\text{CI}}=3$ frames for the CI scenario. Those delay values are picked as they experimentally provide optimal evaluation metrics when comparing the corresponding target to the output of WPE when using the oracle PSD. This setting allows to obtain a real-time factor - defined as the ratio between the time needed to process an utterance and the length of the utterance - below 0.1 with all computations performed on a Nvidia GeForce RTX 2080Ti GPU. A simple decision criterion is used to prevent WPE from updating filter values when the input speech power goes below $-30$~dB, corresponding to speech pauses. Updating the filter with a clean PSD estimated during speech absence indeed provides poor performance as speech resumes.

The DNN used in \cite{Heymann2018} is composed of a single long-short term memory (LSTM) layer with 512 units followed by two linear layers with rectified linear activations (ReLU), and a linear output layer with sigmoid activation. We remove the two ReLU-activated layers in our experiments, as it did not degrade the dereverberation performance, while reducing by $75$~\% the number of trainable parameters.

\subsection{Compared algorithms}

The algorithms evaluated are:
\begin{itemize}
  \setlength\itemsep{-0.20em}
    \item RLS-WPE using the target PSD (\emph{Oracle-WPE})
    \item Classical RLS-WPE (\emph{Vanilla-WPE})\cite{Caroselli2017}
    \item DNN-supported RLS-WPE (\emph{DNN-WPE}) \cite{Heymann2018}
    \item Proposed end-to-end RLS-WPE (\emph{E2E-WPE})
    \item Proposed pretrained E2E-WPE (\emph{E2E-WPE-p})
\end{itemize}

The suffixes \emph{HA} and \emph{CI} correspond to the hearing-aided and cochlear-implanted scenarios, respectively.

\subsection{Evaluation metrics}

We evaluate all approaches on the described test sets. The evaluation is conducted in terms of early-to-late reverberation ratio (ELR) \cite{Carbajal2020b}, perceptual evaluation of speech quality (PESQ), extended short-time objective intelligibility (ESTOI) \cite{Jensen2016} and signal-to-distortion ratio (SDR) \cite{Vincent2006}. The ELR computation uses a separation time of $40$~ms, and is not applicable to evaluating the CI scenario since the target is the direct path only.

\section{Results and Discussion}
\label{sec:results}

We first evaluate the Oracle-WPE approach in the HA scenario, over the first 4 seconds interval and after. As indicated in \tablename~\ref{tab:init}, WPE performance is substantially worse when the filter is not fully initialized. In all further experiments, this initialization period is excluded from evaluation. We then compare the mentioned approaches in the HA scenario (Table~\ref{tab:results-HA}) and the CI scenario (Table~\ref{tab:results-CI}). 

\begin{table}[t]
    \hspace{-0.2cm}
    \scalebox{0.72}{
    \begin{tabular}{c|cccc|cccc|}
        & \multicolumn{4}{c|}{Initialization (4.0 s) } & \multicolumn{4}{c|}{After initialization} \\
        & ELR & PESQ & ESTOI & SDR & ELR & PESQ & ESTOI & SDR \\ 
         \textit{Unprocessed} & \textit{2.9} & \textit{2.29} & \textit{0.61} & \textit{4.0} & \textit{2.9} & \textit{2.26} & \textit{0.61} & \textit{3.9} \\
         Oracle-WPE-HA & 3.0 & 2.49 & 0.65 & 6.5 & \textbf{7.6} & \textbf{2.83} & \textbf{0.77} & \textbf{7.0} \\
    \end{tabular}}
    \caption{\protect\centering Oracle WPE dereverberation performance during and after the initialization period. HA scenario. For all metrics, the higher the better. $\text{T}_\text{60} \in [0.4, 1.0]$}
    \label{tab:init}
\end{table}

\newcommand{\spectrow}{.57\columnwidth}
\newcommand{\spectroxs}{.2\columnwidth}
\newcommand{\spectroys}{0.59\columnwidth}

\begin{figure}[t]
\vspace{-1em}

\begin{tikzpicture}[scale=0.74, transform shape]
 \centering
\begin{axis}
[
    name={clean},
    title = {\textbf{Clean}},
    xmin = 0, xmax = 2,
    ymin = 0, ymax = 8000,
    xtick = {0,1, ...,2},
    ytick = {0,2000,...,8000},
    xlabel = {Time [s]},
    ylabel = {Frequency [Hz]},
    width =\spectrow,
    height =\spectrow
]
\addplot graphics[xmin=0,ymin=0,xmax=2,ymax=8000] {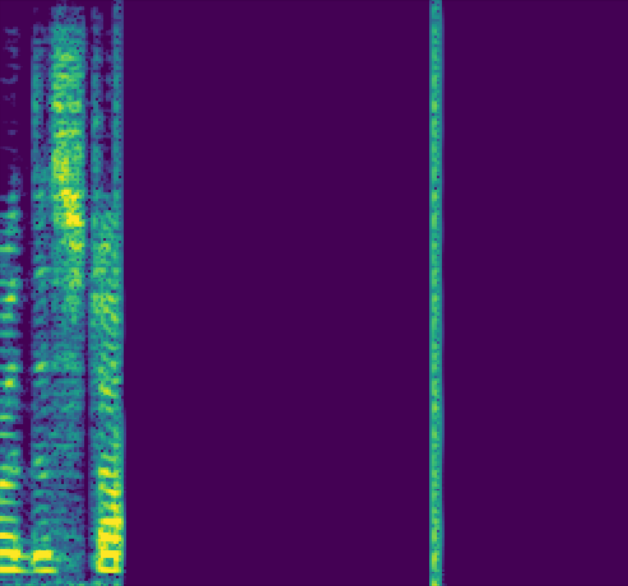};
\end{axis}
 
 \begin{axis}
[
    at={(clean.south east)},
    name={reverberant},
    title = {\textbf{Reverberant}},
    xmin = 0, xmax = 2,
    ymin = 0, ymax = 8000,
    xtick = {0,1, ...,2},
    ytick = {0,2000,...,8000},
    xlabel = {Time [s]},
    ylabel = {Frequency [Hz]},
    width =\spectrow,
    height =\spectrow,
    xshift = \spectroxs
]
\addplot graphics[xmin=0,ymin=0,xmax=2,ymax=8000] {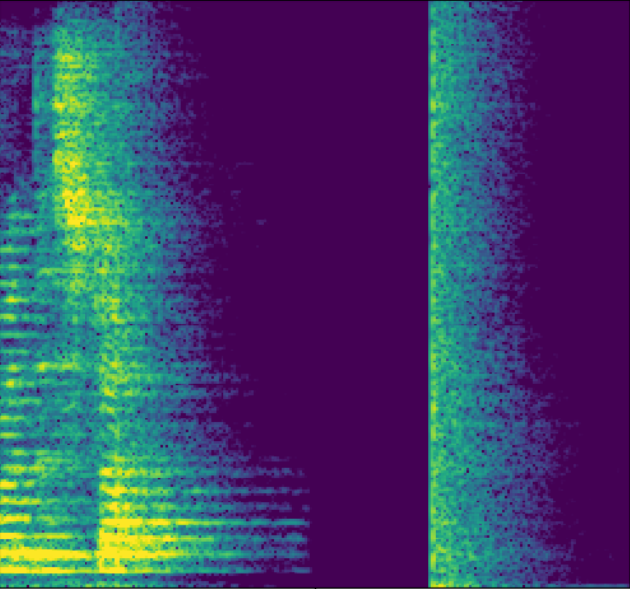};
\end{axis}
  
 \begin{axis}
[
    at={(reverberant.south east)},
    xshift = 0.01\textwidth,
    yshift = -0.25\textwidth,
    width = 0.17\textwidth,
    height = 0.45\textwidth,
    hide axis,
]
\addplot graphics[xmin=0,ymin=0,xmax=1,ymax=1] {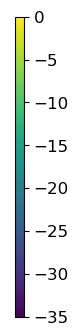};
\end{axis}

 \begin{axis}
[
    at={(clean.south west)},
    yshift=-\spectroys,
    name={wpe},
    title = {\textbf{Vanilla-WPE}},
    xmin = 0, xmax = 2,
    ymin = 0, ymax = 8000,
    xtick = {0,1, ...,2},
    ytick = {0,2000,...,8000},
    xlabel = {Time [s]},
    ylabel = {Frequency [Hz]},
    width =\spectrow,
    height =\spectrow
]
\addplot graphics[xmin=0,ymin=0,xmax=2,ymax=8000] {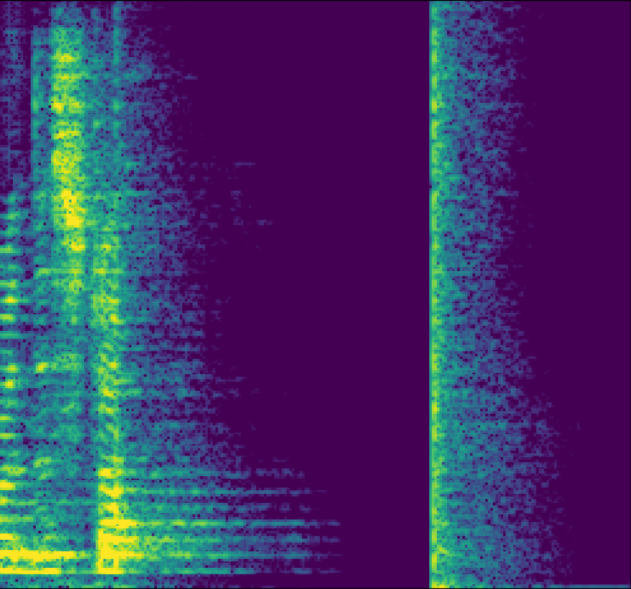};
\end{axis}

  \begin{axis}
[
    at={(wpe.south east)},
    name={E2EWPE},
    title = {\textbf{E2E-WPE-p}},
    xmin = 0, xmax = 2,
    ymin = 0, ymax = 8000,
    xtick = {0,1, ...,2},
    ytick = {0,2000,...,8000},
    xlabel = {Time [s]},
    ylabel = {Frequency [Hz]},
    width =\spectrow,
    height =\spectrow,
    xshift = \spectroxs
]
\addplot graphics[xmin=0,ymin=0,xmax=2,ymax=8000] {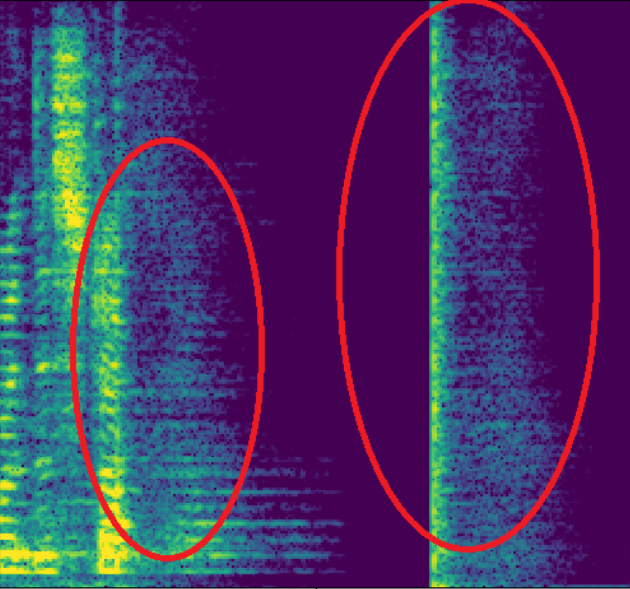};
\end{axis}

\end{tikzpicture}
 
\caption{\protect\centering Log-energy spectrograms of clean, reverberant and processed signals. Dirac impulse following an utterance. $\mathrm{T}_\mathrm{60} = 0.75\mathrm{s}$.}
\label{diracTest}
\end{figure}

\begin{table*}[t]
  \scalebox{0.82}{
  \begin{tabular}{l|cccc|cccc|cccc||cccc|}
     & \multicolumn{4}{c|}{$0.4 \xrightarrow{} 0.6$} & \multicolumn{4}{c|}{$0.6 \xrightarrow{} 0.8$} & \multicolumn{4}{c||}{$0.8 \xrightarrow{} 1.0$} & \multicolumn{4}{c|}{Average} \\
    & ELR & PESQ & ESTOI & SDR & ELR & PESQ & ESTOI & SDR & ELR & PESQ & ESTOI & SDR & ELR & PESQ & ESTOI & SDR \\
    
    \textit{Unprocessed} & \textit{4.9} & \textit{2.49} & \textit{0.70} & \textit{2.8} &      \textit{2.2} & \textit{2.22} & \textit{0.59} & \textit{0.2} &       \textit{0.3} & \textit{2.04} & \textit{0.51} & \textit{-1.6} &     \textit{2.5} & \textit{2.25} & \textit{0.60} & \textit{0.5} \\
    
    \textit{Oracle-WPE-HA} & \textit{11.0} & \textit{3.19} & \textit{0.85} & \textit{5.8} &     \textit{7.0} & \textit{2.77} & \textit{0.77} & \textit{2.8} &     \textit{4.7} & \textit{2.52} & \textit{0.70} & \textit{0.9} &     \textit{7.6} & \textit{2.83} & \textit{0.77} & \textit{3.2} \\
    
    Vanilla-WPE & 11.5 & 3.00 & 0.84 & 6.4 &    8.2 & 2.63 & 0.75 & 4.0 &    6.0 & 2.41 & 0.68 & 2.3 &     8.6 & 2.68 & 0.76 & 4.2 \\
    
    DNN-WPE-HA & 11.3 & 3.06 & 0.85 & 6.1 &    7.5 & 2.67 & 0.76 & 3.4 &     5.1 & 2.43 & 0.69 & 1.5 &    8.0 & 2.72 & 0.77 & 3.7 \\
    
    E2E-WPE-HA & 13.5 & 3.00 & 0.84 & 6.8 &    9.9 & 2.68 & 0.77 & 4.6 &     7.4 & 2.46 & 0.70 & 3.0 &    10.3 & 2.71 & 0.77 & 4.8 \\
    
    E2E-WPE-p-HA & \textbf{13.7} & \textbf{3.07} & \textbf{0.86} & \textbf{6.9} &    \textbf{10.6} & \textbf{2.73} & \textbf{0.78} & \textbf{4.7} &      \textbf{7.8} & \textbf{2.49} & \textbf{0.71} & \textbf{3.1} &    \textbf{10.5} & \textbf{2.76} & \textbf{0.78} & \textbf{4.9} \\

  \end{tabular}}
  \caption{\protect\centering Evaluation results on the HA test set, for different $\text{T}_\text{60}$ reverberation times indicated on the top row in seconds. For all metrics, the higher the better. Best performance is indicated in bold.}
  \label{tab:results-HA}
\end{table*}

\begin{table*}[t]
  \scalebox{0.82}{
  \begin{tabular}{l|cccc|cccc|cccc||cccc|}
     & \multicolumn{4}{c|}{$0.4 \xrightarrow{} 0.6$} & \multicolumn{4}{c|}{$0.6 \xrightarrow{} 0.8$} & \multicolumn{4}{c||}{$0.8 \xrightarrow{} 1.0$} & \multicolumn{4}{c|}{Average} \\
    & ELR & PESQ & ESTOI & SDR & ELR & PESQ & ESTOI & SDR & ELR & PESQ & ESTOI & SDR & ELR & PESQ & ESTOI & SDR \\
    
    \textit{Unprocessed} &   - & \textit{2.29} & \textit{0.58} & \textit{-8.8} &    - & \textit{2.05} & \textit{0.49} & \textit{-10.4} &     - & \textit{1.89} & \textit{0.42}  & \textit{-11.6} &    - & \textit{2.08} & \textit{0.50} & \textit{-10.3} \\
    
    \textit{Oracle-WPE-CI} &  - & \textit{2.91} & \textit{0.76} & \textit{-6.3} &    - & \textit{2.57} & \textit{0.68} & \textit{-8.1} &     - & \textit{2.36} & \textit{0.61} & \textit{-9.3} &    - & \textit{2.61} & \textit{0.68} & \textit{-7.9} \\
    
    Vanilla-WPE &   - & 2.71 & 0.72 & -6.3 &   - & 2.41 & 0.64 & -7.6 &    - & 2.21 & 0.58 & -8.7   & - & 2.44 & 0.65 & -7.6 \\
    
    DNN-WPE-CI &   - & 2.74 & 0.73 & -6.7 &    - & 2.43 & 0.65 & -8.4 &    - & 2.23 & 0.59 & -9.6 &    - & 2.47 & 0.66 & -8.2 \\
    
    E2E-WPE-CI &   - & 2.79 & 0.75 & \textbf{-6.0} &    - & 2.49 & 0.68 & \textbf{-7.4} &    - & 2.28 & 0.62 & \textbf{-8.4} &    - & 2.52 & 0.68 & \textbf{-7.3} \\
    
    E2E-WPE-p-CI &  - & \textbf{2.83} & \textbf{0.76} & -6.2 &   - & \textbf{2.53} & \textbf{0.69} & -7.6 &    - & \textbf{2.32} & \textbf{0.63} & -8.6 &   - & \textbf{2.56} & \textbf{0.69} & -7.4 \\
    
  \end{tabular}}
  \caption{\protect\centering Evaluation results on CI test set, for different $\text{T}_\text{60}$ reverberation times indicated on the top row in seconds. For all metrics, the higher the better. Best performance is indicated in bold.}
  \label{tab:results-CI}
\end{table*}

We notice that for all $\text{T}_\text{60}$ and scenarios, the proposed E2E-WPE-p outperforms its DNN-WPE and Vanilla-WPE counterparts on all metrics. This shows that taking the WPE dereverberation algorithm into account in the DNN optimization process allows the approach to reach an improved output result, without adding any computation nor prior information at test time. We notice that on all metrics except PESQ, the E2E-WPE-p approach performs even slightly better than Oracle-WPE. Our interpretation is that through the end-to-end training procedure, the network does not try to produce an optimal PSD but rather an optimal output. Thus it implicitly modifies the probabilistic nature of the parameter $\lambda_{t,f}$, which then plays the role of a regularizer in \eqref{eq:kalman} rather than that of a variance. Possible explanations are that it either relaxes the Gaussian assumption on the anechoic speech $s$ \cite{Nakatani2008b} or corrects the bias in estimating the time-varying PSD via the periodogram in \eqref{eq:lambda}. As can be seen in \tablename~\ref{tab:results-HA}, using a pretrained DNN significantly helps improving the performance.

Although a filter length of $K=10$ frames and a delay of $\Delta=5$ frames (in the HA scenario) only permits to fully cancel reverberation up to $120$~ms, all approaches achieve significant dereverberation for $\text{T}_\text{60}$ up to $1.0$s. Indeed, the reverberation energy decaying approximately exponentially \cite{Kuttruff2016}, the major part of it resides in the beginning of the reverberation tail. Therefore, although we perceive remains of late reverberation, the objective results are good, especially for the ELR metric which highly reflects this phenomenon.

This contrast between objective improvement and residual reverberation is emphasized with the proposed E2E-WPE(-p) approaches. This is shown in \figurename~2 where an utterance is used to initialize the DNN and WPE statistics and a Dirac impulse is added following 1 second of silence. We notice that the speech contains less short and moderate reverberant energy, yielding a good ELR improvement although some residual late reverberation is present. This is also in line with our informal listening experiments. With the DNN-WPE and Vanilla-WPE approaches, the late reverberation is less identifiable as it is obfuscated by the energy remaining in the short and moderate reverberation through the time-masking phenomenon.

Several approaches to further improve the results may be considered, for instance noise reduction post-processing. As residual late reverberation is perceptually close to noise, it would potentially be a good target for such methods. This is preferred to increasing the prediction filter length of our approach, which results in industrious training while still being unable to cancel very long reverberation.

\section{Conclusion}
\label{sec:conclusion}

We proposed an end-to-end training procedure of the DNN-supported WPE dereverberation algorithm based on \cite{Heymann2018}. The traditional signal processing computations were included into the training of the neural network estimating the anechoic speech PSD. This allowed for specialized training with respect to needs of different listener categories, by letting the network learn customized WPE parameters and targets. 
Results show that this training procedure improved the dereverberation performance without extra computational cost. The approach suppressed most of the reverberation energy immediately following the early reflections, and could be combined with subsequent post-filtering for removing residual late reverberation.

\bibliographystyle{ieeetr}
\bibliography{biblio}

\end{document}